\documentclass[twoside]{dis08}
\usepackage[latin1]{inputenc}
\usepackage[dvips]{graphicx,epsfig,color}
\usepackage{wrapfig,rotating}
\usepackage{amssymb,amsmath,array}

\newcommand{\bea}{\begin{eqnarray}} \newcommand{\eea}{\end{eqnarray}}
\newcommand{\bean}{\begin{eqnarray*}} \newcommand{\eean}{\end{eqnarray*}}
\newcommand{\bm}[1]{\mbox{\boldmath $#1$}}
\newcommand{\s}[1]{{\scriptscriptstyle #1}}
\newcommand{\sT}{{\s T}}
\newcommand{\nn}{\nonumber \\}
\newcommand{\nnn}{\nonumber }

\begin{document}
\title{Spectral Analysis of Gluonic  Pole Matrix Elements}

\author{Leonard Gamberg$^1$\thanks{L.G. acknowledges support from  U.S. Department of Energy under 
contract DE-FG02-07ER41460.} ,  Asmita Mukherjee$^2$, and Piet Mulders$^3$
%
%
\vspace{.3cm}\\
%
1-Penn State University,Berks-Department of Physics\\
Reading,Pennsylvania 19610-USA
%
\vspace{.1cm}\\
2-Indian Institute of Technology-Department of Physics\\
Powai, Mumbai 400076-India
\vspace{.3cm}\\
3-Vrije University-Department of Physics and Astronomy\\
NL-1081 HV Amsterdam, the Netherlands\\
}

\maketitle

\begin{abstract}
We use a spectator framework  to investigate the spectral properties
of quark-quark-gluon correlators and use this to study gluonic pole 
matrix elements. Such matrix elements appear in principle both for 
distribution functions such as the Sivers function and fragmentation 
functions such as the Collins function. 
We find that the contribution of the gluonic pole matrix element in
fragmentation functions vanishes. This outcome is important in the 
study of universality for fragmentation functions.
\end{abstract}

\begin{figure}
\centerline{\includegraphics[width=0.3\columnwidth]{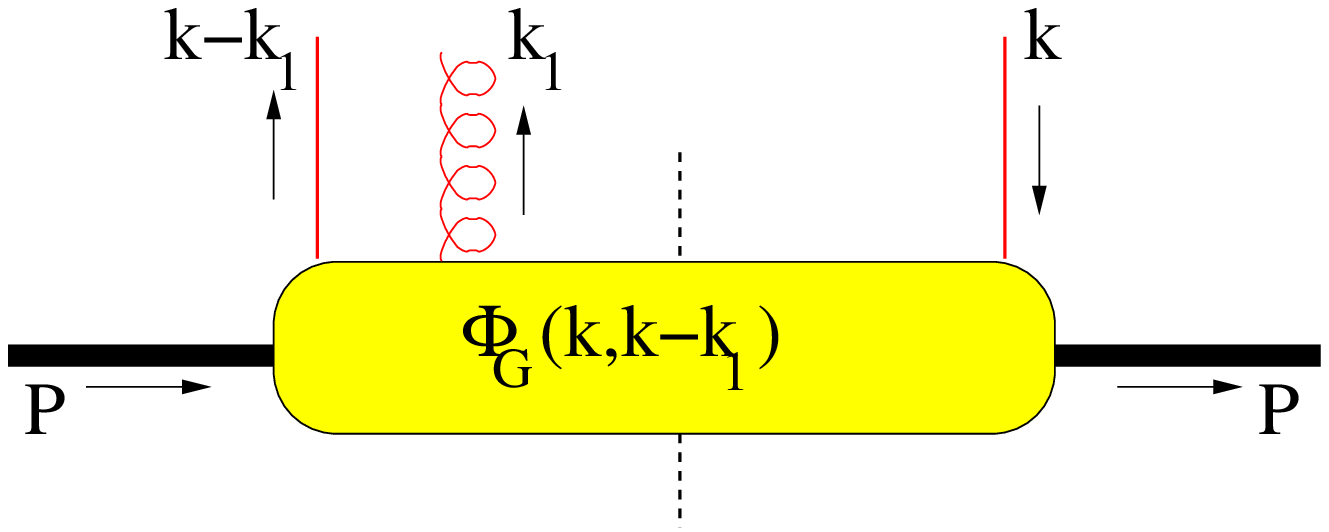}(a)
\includegraphics[width=0.3\columnwidth]{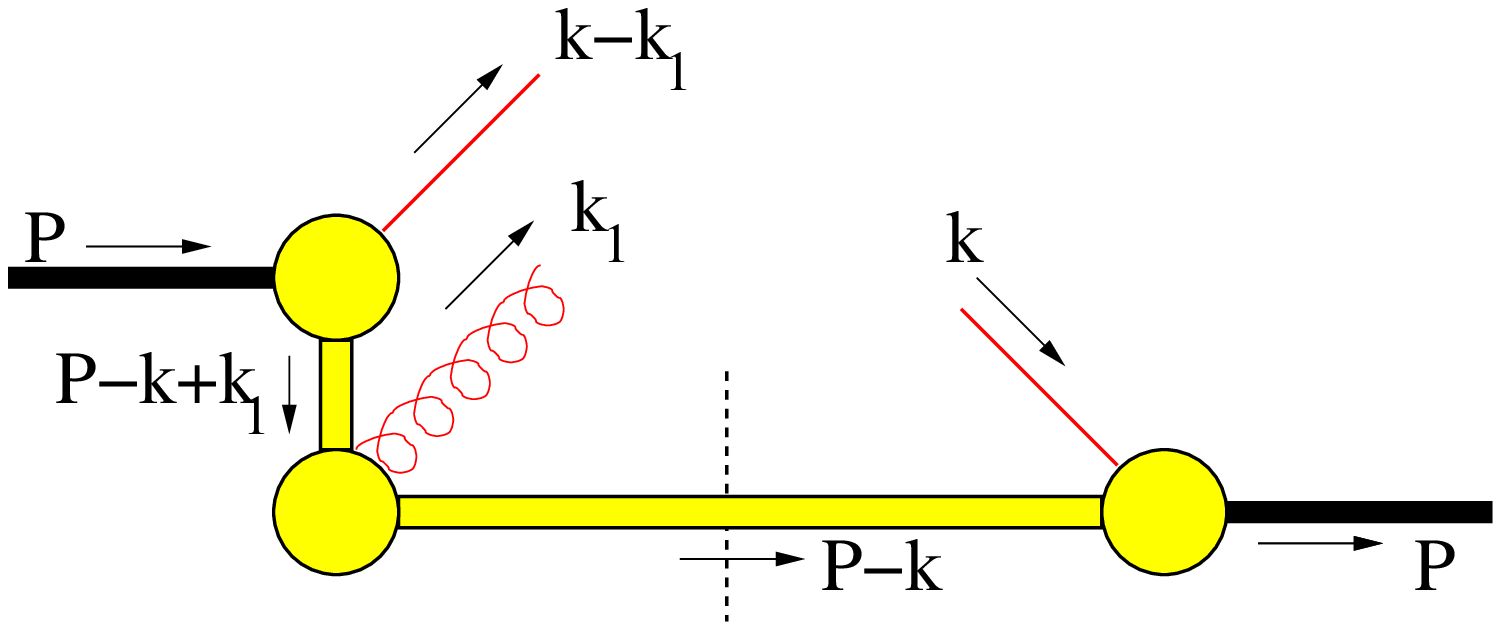}(b)
\includegraphics[width=0.3\columnwidth]{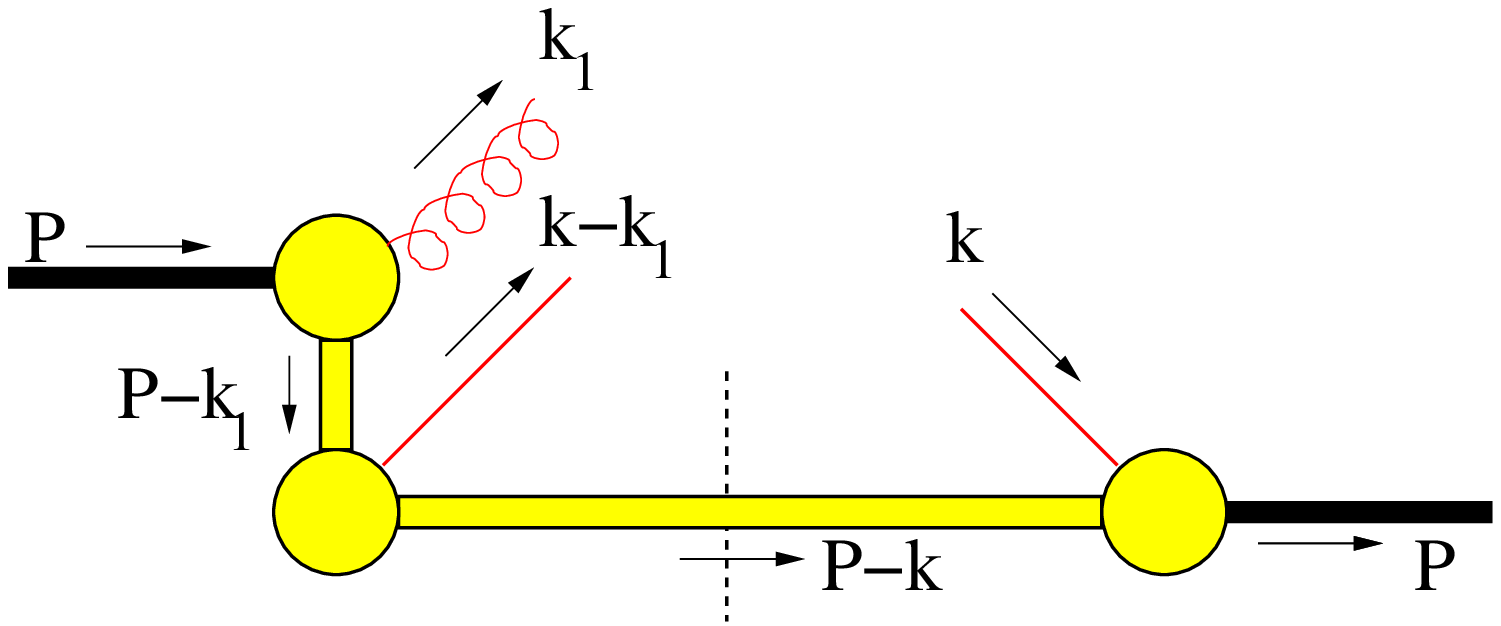}(c)}
\caption{\label{dis}
{ The graphical representation of the quark-quark-gluon
correlator $\Phi_G$ for  the case of distributions including
a gluon with momentum $k_1$ (a), and the
possible intermediate states (b) and (c) 
in a spectator 
model description. Conjugate contribution to (b) and (c)  not shown.}}
\end{figure}

\begin{figure}
\centerline{\includegraphics[width=0.225\columnwidth]{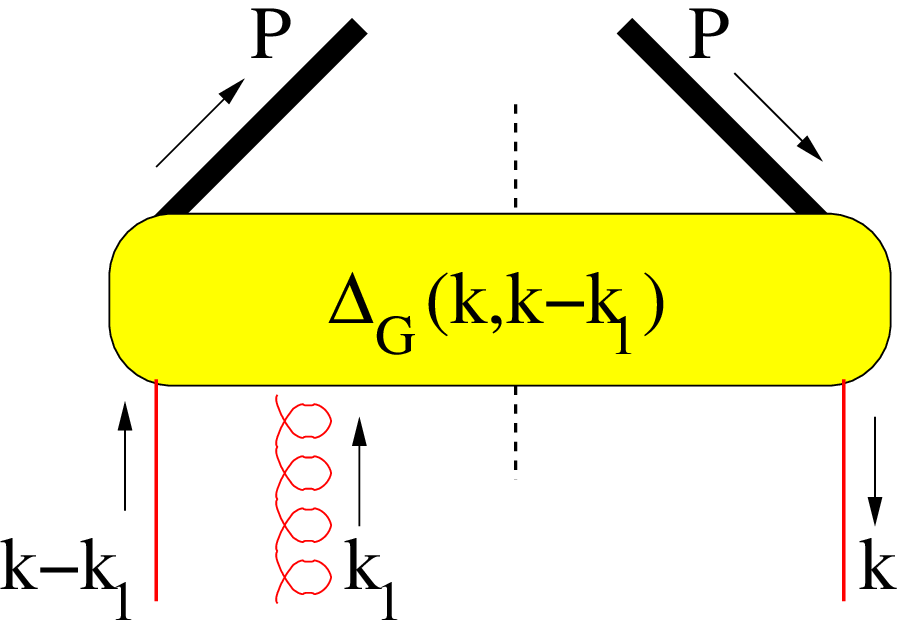}
(a)\includegraphics[width=0.3\columnwidth]{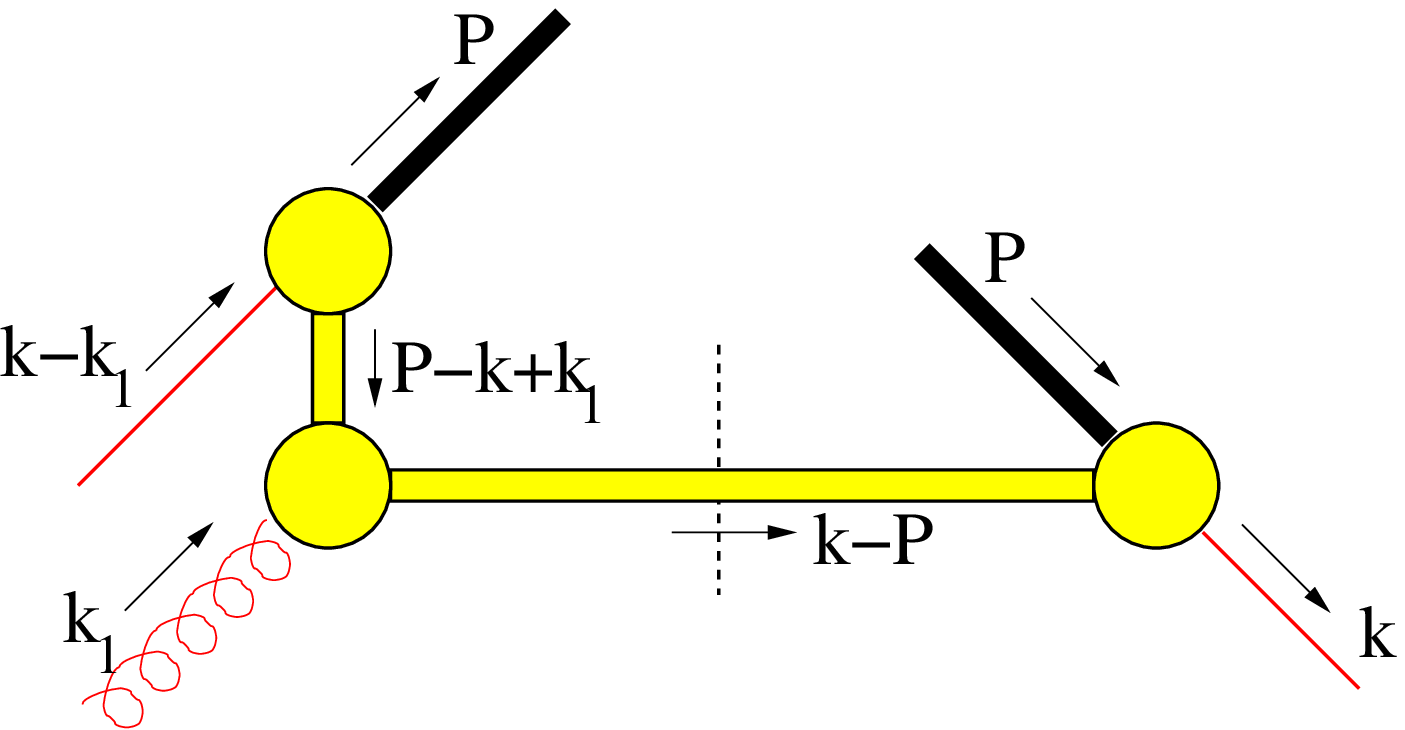}
(b)\includegraphics[width=0.3\columnwidth]{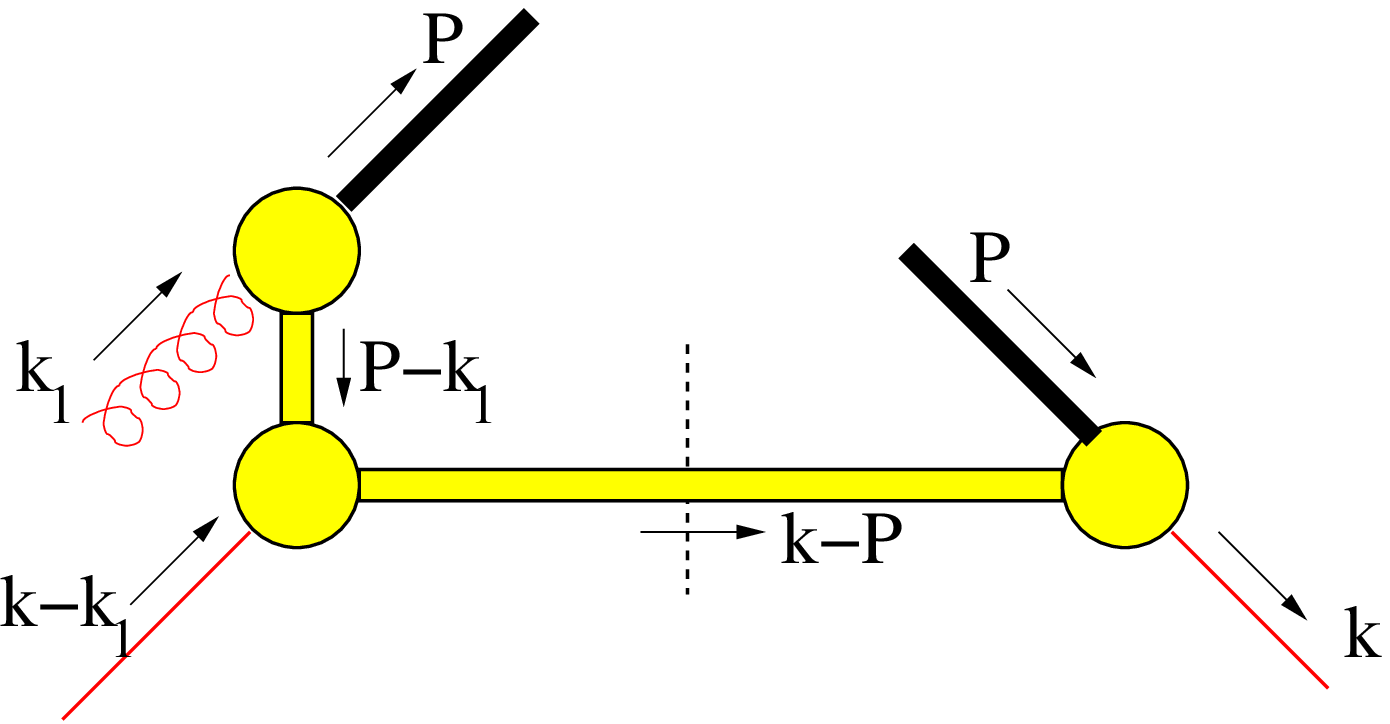}
(c)}
\caption{\label{frag}
{ The graphical representation of the quark-quark-gluon
correlator $\Delta_G$ in the case of fragmentation including
a gluon with momentum $k_1$ (a) and the
possible intermediate states (b)  in a spectator 
model description. Conjugate contribution to (b)  and (c) not shown
}}
\end{figure}

We  investigate multi-parton
correlators with one additional gluon in which the zero-momentum
limit will be studied~\cite{Efremov:1981sh,Qiu:1991pp}. 
These are so-called gluonic pole matrix
elements or Qiu-Sterman matrix elements, that have opposite 
time-reversal (T) behavior as compared to
the matrix elements without the gluon. Such matrix elements involving 
time-reversal odd (T-odd) operator combinations are of interest because
they are essential for understanding single spin asymmetries 
in high energy scattering processes 
e.g. semi-inclusive deep inelastic scattering (SIDIS) 
and Drell-Yan scattering.   In order
to understand the basic features of these matrix elements we 
perform a spectral analysis by modeling the distribution and
fragmentation functions under {\em reasonable} 
assumptions~\cite{Gamberg:2008yt}.  
In particular we consider the differences between distribution
and fragmentation functions using a spectral analysis  
while restricting the momentum dependence and
asymptotic behavior of the vertices.   
In this context, the relevant gluonic pole matrix elements 
that we want to  study are 
$\Phi_G(k,k-k_1)$ and $\Delta_G(k,k-k_1)$ shown in 
Figs.~\ref{dis} and \ref{frag}. Of these matrix
elements only the dependence on the collinear components
$x$ and $x_1$ in the expansion of the momenta is needed
(note, the gluon momentum is parameterized as
$k_1 =[k_1^-,x_1, k_{1\sT}]$ in these figures).
We find that while both $\Phi_G(x,x-x_1)$
and $\Delta_G(x,x-x_1)$ are nonzero,
taking the limit $x_1 \rightarrow x$, $\Phi_G(x,x)$ 
remains non-zero, while $\Delta_G(x,x)$ vanishes.  
The vanishing of the T-odd gluonic pole matrix elements is important
in the study of universality of 
transverse momentum dependent (TMD) distribution and 
fragmentation functions (FFs). 

T-odd {\em operator structure}
can be traced to the color gauge link that necessarily appears in
correlators to render them color gauge-invariant~\cite{Belitsky:2002sm,Boer:2003cm}. The quark-quark correlator
depending on the collinear and transverse components
of the quark momentum, $k=x\,P+\sigma\,n+k_\sT$ (where the
Sudakov vector $n$ is an arbitrary light-like four-vector $n^2=0$ 
that has non-zero overlap $P\cdot n$ with the hadron's momentum $P$ and 
$k^-\sim\sigma$ which is suppressed w/r to the hard scale) 
 is given by,
\bea\label{TMDDF}
\Phi_{ij}^{[\mathcal U]}(x{,}k_\sT)
={\int}\frac{d(\xi{\cdot}P)\,d^2\xi_\sT}{(2\pi)^3}\ e^{ik\cdot\xi}
\langle P|\,\overline\psi_j(0)\,\mathcal U_{[0;\xi]}\,
\psi_i(\xi)\,|P\rangle\big\rfloor_{\text{LF}}\ ,
\eea
where LF ($\xi\cdot n=0$) designates the light-front. 
The  \emph{gauge link} is the path-ordered exponential, 
$\mathcal U_{[\eta;\xi]}
=\mathcal P{\exp}\big[{-}ig{\int_C}\,ds{\cdot}A^a(s)\,t^a\,\big]$
 along the integration path $C$ with
endpoints at $\eta$ and $\xi$.
Its presence in the hadronic matrix element is required by gauge-invariance.
Similarly, the fragmentation 
correlator depending on the collinear and transverse components of the
quark momentum, $k = \tfrac{1}{z}\,P + k_\sT + \sigma\,n$,
is given by~\cite{Boer:2003cm}
\bea
\Delta^{[\mathcal U]}_{ij}(z,k_\sT)&=&
\sum_X\int\frac{d(\xi\cdot P_h)\,d^2\xi_\sT}{(2\pi)^3}\ e^{i\,k\cdot\xi}
\langle 0 |\mathcal U_{[0,\xi]}\psi_i(\xi)|P,X\rangle
\langle P,X|\bar{\psi}_j(0)|0\rangle |_{LF}\,  .
\label{TMDFF}
\eea
In the correlators the integration path $C$ in
the gauge link designates  process-dependence. 
This is due to the observation
that the operator
structure of the correlator is also a consequence of the necessary
resummation of all contributions that arise from collinear gluon
polarizations, i.e.\ those along the hadron momentum. How this 
resummation takes effect is a matter of calculation~\cite{Bomhof:2006ra}. 
The result is a process dependence in the path in the gauge link. 
After azimuthal weighting of cross sections one 
simply finds that the T-odd features originating from the gauge link
lead to specific factors with which the T-odd functions appear in 
observables. For example, comparing T-odd effects in DFs in
SIDIS and the Drell-Yan 
process one finds a relative minus sign~\cite{Collins:2002kn,Brodsky:2002rv}. 
Similarly, comparing T-odd 
effects in FFs in SIDIS and electron-positron
annihilation one also finds a relative minus sign  for the
T-odd effect originating from the operator 
structure (gauge link)~\cite{Boer:2003cm}.
The T-odd operator parts
are precisely the soft limits ($k_1\rightarrow 0$ or $x_1$ and
$z^-1=x_1\rightarrow 0$) 
of the gluonic pole matrix elements~\cite{Boer:2003cm} 
\bea
\Phi_\partial^{\alpha\,[\mathcal U]}(x)
=\tilde \Phi_\partial^{\alpha}(x)
+C_G^{[\mathcal U]}\,\pi\Phi_G^{\alpha}(x,x)\, , \quad {\rm and}
\quad\Delta^{\alpha\,[\mathcal U]}_{\partial}(z)
=\tilde{\Delta}_\partial^\alpha(\frac{1}{z})
+C_G^{[\mathcal U]}
\,\pi\Delta_G^\alpha(\frac{1}{z},\frac{1}{z})\, 
\label{decomp}
\eea
(see Figs.~\ref{dis} and \ref{frag}).  
They arise in the decomposition of   the transverse weighted quark
 correlators
\bea
\label{TransverseMoment}
\Phi_{\partial}^{\alpha\,[\mathcal U]}(x) 
= \int d^2k_\sT\ k_\sT^\alpha\,\Phi^{[\mathcal U]}(x{,}k_\sT)\, ,\quad 
{\rm and} \quad
\Delta^{\alpha\,[\mathcal U]}_{\partial}(z)
=\int d^2k_\sT\ k_\sT^\alpha \Delta^{[\mathcal U]}(z,k_\sT)\, ,
\eea
which are the relevant operators in analyzing the azimuthal asymmetries. 
The process-dependent gluonic pole factors $C_G^{[\mathcal U]}$ are calculable
and  the process (link) independent correlators $\tilde\Phi_\partial$ and
 $\tilde{\Delta}_\partial$ contains the T-even operator 
combination, while $\Phi_G$ and $\Delta_G$ contain the T-odd operator combination.  The latter one is 
precisely the soft limit, $z_1^{-1} = x_1 \rightarrow 0$, 
of the quark-gluon correlator 
$\Delta_{G\,ij}^\alpha(x,x_1)$
\bea
\Delta_{G\,ij}^\alpha\left(x,x-x_1\right)
&=&\sum_X \int\frac{d(\xi{\cdot}P)}{2\pi}\frac{d(\eta{\cdot}P)}{2\pi}\,
e^{i\,x_1(\eta\cdot P)}e^{i\,(x-x_1)(\xi\cdot P)}\,
\nn &&\times
\langle 0 | \mathcal U^n_{[0,\eta]}\, gG^{n\alpha}(\eta)
\,\mathcal U^n_{[\eta,\xi]}\psi_i(\xi)|P,X\rangle
\langle P,X|\overline{\psi}_j(0)|0\rangle\Bigg|_{LC} .
\nnn
\label{GLa}
 \eea
Because of the appearance of hadronic states $\vert P,X\rangle$,
each of correlators 
in $\quad\Delta^{\alpha\,[\mathcal U]}_{\partial}(z)$ 
contains in principle 
T-even and T-odd functions. However, 
rather than having a doubling of T-odd functions, 
we find that $\Delta_G(x,x)$ = 0,
which implies that T-odd fragmentation functions
appear in the matrix elements
of the T-even operator combination in $\tilde \Delta_\partial$ involving
a hadron-jet state (non-plane-wave). 
They are process independent 
for instance the T-odd Collins function~\cite{Collins:1992kk} and
they  appear with a universal strength (no gluonic pole factors).  
In contrast T-odd DFs in
$\Phi_\partial$ only can come from $\Phi_G(k_1=0)$. These DFs can still
be universal but appear with calculable process dependent gluonic pole 
factors~\cite{Bomhof:2004aw,Bomhof:2006ra}.

To see this in 
a spectator model approach, we consider
the  distribution or fragmentation
correlators with a spectator with mass $M_s$. The 
result for the cut, but untruncated, diagrams, such as in 
Figs.~\ref{dis} and \ref{frag} (without the gluon insertion) 
are of the form
\bea
\Phi(x,k_\sT)\hspace{-.2cm} &\sim &\hspace{-.2cm}\int d(k\cdot P)
\frac{F(k^2,k\cdot P)}{(k^2-m^2+i\epsilon)^2}
\delta\left( (k-P)^2 - M_s^2\right),
\nnn
\label{basic}
\eea
where  $F(k^2,k\cdot P)$  contains the numerators of propagators and/or
traces of them in the presence of Dirac Gamma matrices, as well as 
the vertex form factors (see for example~\cite{Jakob:1997wg}).
In the above the delta function constraint in Eq.~\ref{basic} has
been implemented. One finds that the numerator
$F(k^2,k\cdot P) = F(x,k_\sT^2)$ and hence
\bea
\Phi(x,k_\sT) \sim 
\ \frac{(1-x)^2\,F(x,k_\sT)}{\left(\mu^2(x)-k_\sT^2\right)^2},
\label{qqspec}
\eea
with $\mu^2(x) = x\,M_s^2+(1-x)\,m^2-x(1-x)\,M^2.$
Note that $k_\sT^2 = -\bm k_\sT^2 \le 0$.
The details of the numerator function depend on the details of the 
model, including the vertices, polarization sums, etc. These must
be chosen in such a way as to not produce unphysical effects,
such as a decaying proton if $M \ge m+M_s$, thus $m$ in Eq.~\ref{basic}
must represent some constituent mass in the quark propagator, rather
than the bare mass.
The useful feature of the result in Eq.~\ref{qqspec} is its ability to
produce reasonable valence and even sea quark distributions using the
freedom in the model.
The results for the fragmentation function in the spectator
model is identical upon the substitution of $x = 1/z$~\cite{Jakob:1997wg}. 

We now turn to the same spectral analysis of the gluonic pole
correlator using the picture given in Figs.~\ref{dis} and~\ref{frag} for
distribution and fragmentation functions respectively.
 Again, we only need to investigate
one of the cases. Parameterizing 
the gluon momentum as $k_1 = [k_1^-,x_1, k_{1\sT}]$,
 $k_1^- = k_1\cdot P - \tfrac{1}{2}\,x_1\,M^2$ 
is the first component to be integrated over~\cite{Gamberg:2008yt}.
Assuming that the numerator does not grow with $k_1^-$ one can
easily perform the $k_1^-$ integrations 
assuming that the $F_i$ are independent of $k_1^-$.  
Taking the limit $x_1 \rightarrow 0$ 
of the basic result for the quark-gluon correlators 
$\Phi_G(x,x-x_1,k_\sT,k_\sT-k_{1\sT})$
we obtain  the gluonic pole correlators,
for distribution functions 
($0\le x \le 1$) (see~\cite{Gamberg:2008yt} for details)
\bea
\Phi_G(x,x) 
&=& - \int d^2k_\sT\,d^2k_{1\sT}
\ \frac{(1-x)\,F_1(x,0,k_\sT,k_{1\sT})\theta(1-x)}{
\bigl(\mu^2-k_\sT^2\bigr) \bigl(x\,B_1+(1-x)\,A_2\bigr)\,A_1}\, ,
\eea
where $A_i(\{m_i^2\},\{k_{iT}^2\},\{x_i\})$,  
and for fragmentation functions ($x = 1/z \ge 1$) 
\bea
\Delta_G(x,x) &=& 0\, .
\eea
This result depends on the assumption that the numerator
does not grow with $k_1^-$, otherwise,
 one does not get the required $x_1\,\theta(x_1)$ behavior in the
calculation~\cite{Gamberg:2008yt}. 
In  models, terms proportional to $k_1^- \sim k_1\cdot P$ 
may easily arise from numerators of fermionic 
propagators~\cite{Gamberg:2006ru}
which may easily be suppressed by  form factors at the
vertices. To prove a proper behavior within QCD one would need to study the
fully unintegrated correlators such as e.g.\ in 
Ref.~\cite{Collins:2007ph}
and show that they fall off sufficiently fast as a function of $k_1\cdot P$.
 
While our analysis is  not yet the full proof that 
gluonic pole  matrix elements vanish in the case of fragmentation,
it is a step towards such a  proof and 
the possible direction to
obtain such a proof by  considering the appropriate color gauge-invariant
soft matrix elements.
Such a proof is important as it eliminates a whole class of matrix
elements parameterized in terms of T-odd fragmentation functions besides
the T-odd fragmentation functions 
in the parameterization of the two-parton correlators.


\begin{footnotesize}


\bibliographystyle{unsrt}
\bibliography{gamberg_leonard_refs}
\end{footnotesize}


\end{document}